\documentclass[prb,twocolumn,superscriptaddress,showpacs,amsmath,amssymb]{revtex4}
\usepackage{amsfonts}
\usepackage{bbm}
\usepackage{color}
\usepackage{graphicx}
\usepackage{dcolumn}
\usepackage{bm}

\begin{document}
\title{Spin-triplet superconductor$-$quantum anomalous Hall insulator$-$spin-triplet superconductor Josephson junctions: $0$-$\pi$ transition, $\phi_{0}$ phase and switch effects}

\author{Qiang Cheng}
\affiliation{School of Science, Qingdao University of Technology, Qingdao, Shandong 266520, China}
\affiliation{International Center for Quantum Materials, School of Physics, Peking University, Beijing 100871, China}

\author{Qing Yan}
\affiliation{International Center for Quantum Materials, School of Physics, Peking University, Beijing 100871, China}
\affiliation{Collaborative Innovation Center of Quantum Matter, Beijing 100871, China}

\author{Qing-Feng Sun}
\email[]{sunqf@pku.edu.cn}
\affiliation{International Center for Quantum Materials, School of Physics, Peking University, Beijing 100871, China}
\affiliation{Collaborative Innovation Center of Quantum Matter, Beijing 100871, China}
\affiliation{CAS Center for Excellence in Topological Quantum Computation, University of Chinese Academy of Sciences, Beijing 100190, China}

\begin{abstract}
We study the Josephson effect in spin-triplet superconductor$-$quantum anomalous Hall insulator$-$spin-triplet superconductor junctions using the nonequilibrium Green function method.
The current-phase difference relations show strong dependence on the orientations of the $\bf{d}$-vectors in superconductors.
We focus on two $\bf{d}$-vector configurations,
the parallel one with the left and right ${\bf{d}}$-vectors
being in the same direction, and the nonparallel one
with the left ${\bf{d}}$-vector fixed at the $z$-axis.
For the parallel configuration, the $0$-$\pi$ transition can be realized when one rotates the ${\bf{d}}$-vectors from the parallel to the junction plane to the perpendicular direction. The $\phi_{0}$ phase with nonzero Josephson current at zero phase difference can be obtained as long as ${d_{x}}{d_{z}}\ne0$.
For the nonparallel configuration, the $0$-$\pi$ transition and the $\phi_{0}$ phase still exist. The condition for the formation of the $\phi_{0}$ phase becomes $d_{Rx}\ne0$.
The switch effects of the Josephson current are found in both configurations when the ${\bf{d}}$-vectors are rotated in the $xy$ plane.
Furthermore, the symmetries satisfied by the current-phase difference
relations are analysed in details by the operations of the time-reversal, mirror-reflections, the spin-rotation and the gauge transformation,
which can well explain the above selection rules for the $\phi_{0}$ phase.
Our results reveal the peculiar Josephson effect between spin-triplet superconductors and the quantum anomalous Hall insulator, which provide helpful phases and effects for the device designs. The distinct current-phase difference relations for different orientations may be used to determine the direction of the ${\bf{d}}$-vector in the spin-triplet superconductor.

\end{abstract}
\maketitle

\section{\label{sec1}Introduction}
The quantum anomalous Hall insulator (QAHI) with bulk gap and chiral edge states in the absence of external magnetic field has been experimentally observed in the magnetic topological insulator\cite{Chang} soon after its theoretical prediction\cite{Yu}. QAHI can realize the chiral topological superconducting states when it is in proximity to a conventional $s$-wave superconductor\cite{Qi}. Various studies on the electrical transport have been carried out to detect or regulate the chiral Majorana edge modes produced in the composite system\cite{Wang,Zhou,He,Lian,Chen,Yan,Zhang,Huang,Ii}. The Josephson junctions are also researched, which exhibit novel phase shift\cite{Sakurai,Yan2}, anomalous critical current\cite{Chen2}, tunable Majorana valve effect\cite{Li} or induced paring states\cite{Nakai}. However, the superconductors involved in the existing studies are limited to the spin-singlet pairing. The form of interaction between QAHI and the spin-triplet superconductors (STSs) is still unknown.

Generally, STSs show more physics due to thier complex spin structures of Cooper pairs\cite{Mackenzie}. The spin part of the superconducting wave function is described by the so called ${\bf{d}}$-vector which has three components in a rectangular coordinate system, i.e., ${\bf{d}}=(d_{x},d_{y},d_{z})$.\cite{Balian} Its direction can be tuned by a very weak field\cite{Annett}. The orientation of the ${\bf{d}}$-vector can impose decisive impact on the transport and topological properties of STS\cite{Terrade,Mercaldo,Brydon}. Especially, for the magnetic Josephson junctions, the relative orientation of two ${\bf{d}}$-vectors in STSs can be used to adjust the Andreev bound states\cite{Mercaldo2} and to produce the Josephson current switches\cite{Kastening} or the $0$-$\pi$ phase transitions\cite{Brydon2,Cheng} valuable for the circuit element of quantum computation\cite{Gingrich}. For the material realization of STSs, there are many theoretical and experimental researches for the identification of the spin-triplet pairing\cite{Mackenzie,Metz,Ghosh,Shang}, which include the determination of the direction of the ${\bf{d}}$-vector\cite{KTanaka,Kaladzhyan}.
In addition, the spin-triplet pairing with a nonzero ${\bf{d}}$-vector also appears
in some superconducting material with the spin-orbit coupling.\cite{addzhou,addlv}
In this paper, we study the STS-QAHI-STS Josephson junctions with the chiral $p$-wave pairing in STSs. The ${\bf{d}}$-vectors are expressed as $(k_{x}+i k_{y})$ for their orbital part. This type of paring is believed to be the candidate state for $\text{Sr}_{2}\text{RO}_{4}$.\cite{Mackenzie,Ikegaya}

In our STS-QAHI-STS junctions, the two ${\bf{d}}$-vectors in STSs can be along any directions. For definiteness, we study the current-phase difference relations (CPRs) for two configurations using the lattice nonequilibrium Green function technology. For the first configuration, the vectors keep parallel and are rotated simultaneously. We find if the orientation of ${\bf{d}}$-vectors is changed from the direction parallel to the junctions to that perpendicular to the junctions, the $0$-$\pi$ transition will happen. When the ${\bf{d}}$-vectors satisfy the condition ${{d}}_{x}{{d}_{z}}\ne0$, the $\cos{\phi}$-type current emerges. The $\phi_{0}$ phase with free energy minimum at the phase difference $\phi\ne0,\pi$ forms. This phase possesses the nonzero current as the phase difference $\phi$ is zero, which has attracted numerous theoretical and experimental researches\cite{Szombati,Yokoyama,Alidoust,Dolcini,Liu,Buzdin} due to its potential applications in device designs\cite{Padurariu}.
For the second configuration, the ${\bf{d}}$-vector for the left STS is fixed along the $z$-axis, while that for the right STS is rotated arbitrarily. It is found the $0$-$\pi$ transition happens when the right vector is inverted from the $+z$ direction to the $-z$ direction. When the $x$ component of the right vector is not zero, i.e., $d_{Rx}\ne0$, the $\cos{\phi}$-type current appears and the $\phi_{0}$ phase forms. We also find the on/off effects of the Josephson current for both configurations when the ${\bf{d}}$-vectors are rotated from the $x$ direction to the $y$ direction in the $xy$ plane.

In addition, three universal symmetry relations for CPRs in STS-QAHI-STS junctions are derived, which apply to the general ${\bf{d}}$-vector configuration. These relations can well explain the novel behaviours of CPRs including the selection rules for the $\cos{\phi}$-type current and the $\phi_{0}$ phase. To clarify the origin of the relations, we analyse the invariance of QAHI using the continuum model under operations of the time-reversal, mirror-reflections, the spin-rotation and the gauge transformation, as well as the changes imposed on STSs by the operations.
From the analyses, we find the symmetry relations actually reflect the unique nature of QAHI and its peculiar interaction with STSs.

The rest of paper is organized as follows. In Sec. \uppercase\expandafter{\romannumeral2}, we present the continuum and lattice models for QAHI and STSs. The edge states of QAHI are solved with the continuum Hamiltonian. The Josephson current is expressed based on the lattice model by the nonequilibrium Green function method. In Sec. \uppercase\expandafter{\romannumeral3}, the numerical results are presented for the parallel and nonparallel configurations of ${\bf{d}}$-vectors. The $0$-$\pi$ transition, the selection rules for the $\phi_{0}$ phase and the symmetry relations for CPRs are discussed in detail. Sec. \uppercase\expandafter{\romannumeral4} analyses the origin of the symmetry relations through the continuum models under five kinds of transformations.
At last, the results are summarized in Sec.\uppercase\expandafter{\romannumeral5}.

\section{\label{sec2}Model and formulation}
\subsection{\label{sec2.1}Continuum model}
\begin{figure}[!htb]
\centerline{\includegraphics[width=1\columnwidth]{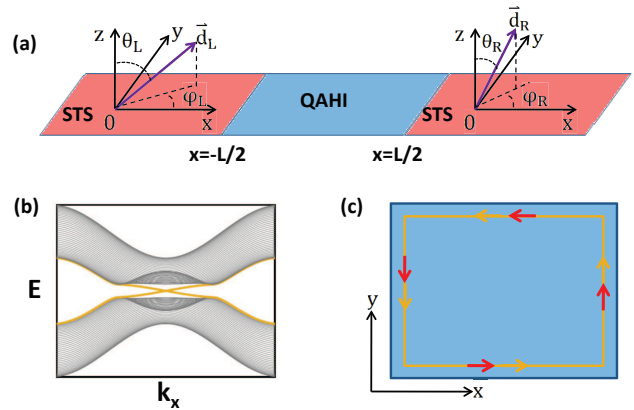}}
\caption{(a) Schematic illustration of the STS-QAHI-STS junctions. The $\bf{d}$-vectors in the left and the right STSs are denoted by $\bf{d}_L$ and $\bf{d}_R$, respectively. The direction of $\bf{d}_L$($\bf{d}_R$) is depicted by the polar angle $\theta_L$($\theta_R$) and the azimuthal angle $\varphi_L$($\varphi_R$). The junctions are placed in the $xy$ plane. (b) The energy bands of QAHI. The linear dispersions (yellow) for the edge states are located in the gap of bulk bands(grey). (c) The edge states in QAHI. The yellow arrows denote the motion direction of electrons and the red arrows denote their spin.}
\label{fig1}
\end{figure}

We consider the two-dimensional STS-QAHI-STS Josephson junction in the $xy$ plane as shown in Fig. 1(a). The finite width along the $y$ direction of the junctions is $W$. The length of QAHI is $L$ and is limited in the region $-\frac{L}{2}<x<\frac{L}{2}$. The semi-infinite STSs are placed in the region $x<-\frac{L}{2}$ and $x>\frac{L}{2}$ for the left one and the right one, respectively.
 The Hamiltonian of the junctions is written as
\begin{eqnarray}
H=H_{LS}+H_{QAHI}+H_{RS},\label{HLQR}
\end{eqnarray}
where $H_{LS}$, $H_{QAHI}$ and $H_{RS}$ are the Hamiltonians for the left STS, QAHI and the right STS, respectively.

For the continuum model, the STS Hamiltonian $H_{L(R)S}$ is given by (we take $\hbar=1$.)
\begin{eqnarray}
H_{L(R)S}=\sum_{\bf{k}}\Psi^{\dagger}_{L(R)\bf{k}}\check{H}_{L(R)}({\bf{k}})\Psi_{L(R)\bf{k}},\label{HL}
\end{eqnarray}
with $\Psi_{L(R)\bf{k}}=(c_{L(R),\bf{k}\uparrow},c_{L(R),\bf{k}\downarrow},c^{\dagger}_{L(R),-\bf{k}\uparrow},c^{\dagger}_{L(R),-\bf{k}\downarrow})^{T}$ and the $4\times4$ Bogoliubov-de Gennes (BdG) Hamiltonian
\begin{eqnarray}
\check{H}_{L(R)}({\bf{k}})=
\left(\begin{array}{cc}
\epsilon_{L(R)}&({\bf{\sigma}}\cdot{\bf{d}}_{L(R)})i\sigma_y\\
({\bf{\sigma}}\cdot{\bf{d}}_{L(R)})^{*}i\sigma_y&-\epsilon_{L(R)}
\end{array}\right).
\end{eqnarray}
Here $\epsilon_{L(R)}=\frac{k^2}{2m}-\mu_{L(R)}$, ${\bf{d}}_{L(R)}=\Delta f({\bf{k}})e^{i\phi_{L(R)}}{\bf{n}}_{L(R)}$ and the Pauli matrices ${\bf{\sigma}}=(\sigma_x,\sigma_y,\sigma_z)$. The chemical potential and the energy gap are denoted by $\mu_{L(R)}$ and $\Delta$, respectively. We choose the chiral $p$-wave pairing with $f({\bf{k}})=k_{x}+i k_y$ for STSs\cite{Mackenzie,Ikegaya}, in which ${\bf{k}}=(k_x,k_y)$ is the two-dimensional wavevector. The direction of ${\bf{d}}_{L(R)}$ is expressed by its polar angle $\theta_{L(R)}$ and azimuthal angle $\varphi_{L(R)}$, i.e., ${\bf{n}}_{L(R)}=(n_{L(R)1},n_{L(R)2},n_{L(R)3})
=(\sin\theta_{L(R)}\cos{\varphi_{L(R)}},\sin\theta_{L(R)}\sin{\varphi_{L(R)}},\cos{\theta_{L(R)}})$, as shown in Fig. 1(a).

For QAHI, we adopt the following Hamiltonian of the BdG form\cite{Yu},
\begin{eqnarray}
H_{QAHI}=\sum_{\bf{k}}{\psi}^{\dagger}_{\bf{k}}\check{H}_{QAHI}({\bf{k}}){\psi}_{\bf{k}},\label{HQ}
\end{eqnarray}
with ${\psi}_{\bf{k}}=(c_{\bf{k}\uparrow},c_{\bf{k}\downarrow},c^{\dagger}_{-\bf{k}\uparrow},c^{\dagger}_{-\bf{k}\downarrow})^{T}$ and
\begin{eqnarray}
\check{H}_{QAHI}({\bf{k}})=\left(\begin{array}{cc}
h({\bf{k}})&0\\
0&-h^{*}(-\bf{k})\end{array}\right).
\end{eqnarray}
The electron part is given by
\begin{eqnarray}
h({\bf{k}})=\left(\begin{array}{cc}
m_0+Bk^2&A(k_x-i k_y)\\
A(k_x+i k_y)&-m_0-Bk^2
\end{array}\right),\label{QE}
\end{eqnarray}
where the parameters are taken as $m_0=-1.5$ and $A=B=0.25$ in this paper. This will lead to the energy bands as shown in Fig. 1(b) for QAHI with the periodic boundary condition in the $x$ direction and open boundary conditions at $y=0$ and $W$. There is a bulk energy gap for electrons and linear dispersions at edges. The linear dispersion with the positive slope corresponds to the edge $y=0$ and the negative slope corresponds to the edge $y=W$.

From the continuum model Eq.($\ref{QE}$), we can solve the edge states for energy $E>0$ at $y=0$. The dispersion relation is
\begin{eqnarray}
E=Ak_{x},\label{ed}
\end{eqnarray}
and the wave function is
\begin{eqnarray}
c\left(\begin{array}{c}
1\\
1\end{array}\right)e^{-\frac{A}{2B}y}\sin{\big[y\sqrt{-k^2_{x}-\frac{m_{0}}{B}-\frac{A^2}{4B^2}}\big]},\label{ewf}
\end{eqnarray}
with a constant $c$. The wave function decays towards the interior of QAHI and the spin of the electron is along the $x$-axis. The edge state at $y=W$ can be solved in a similar way. If QAHI is also finite along the $x$ direction such as the situation for the junctions in Fig. 1(a), there will be four edges. The edge sates are plotted in Fig. 1(c). The yellow arrows denote the motion direction of electrons and red arrows represent their spin.

\subsection{\label{sec2.2}Lattice model}
\begin{figure}[!htb]
\centerline{\includegraphics[width=1\columnwidth]{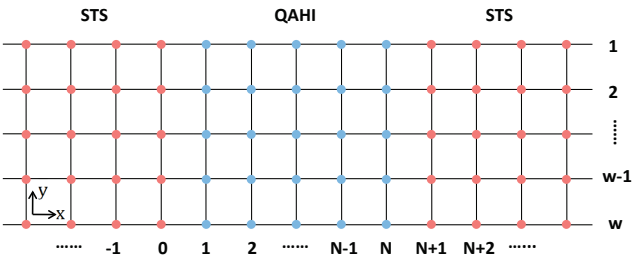}}
\caption{Schematic illustration of two-dimensional square lattice model for the Josephson junctions in Fig. 1(a). The lattice constant is $a$. The width of the lattice is $w$. The length of QAHI is $N$. In our calculations, we take $N=40$ and $w=40$.}
\label{fig1}
\end{figure}
In order to calculate the Josephson current, we discretize the continuum Hamiltonians on a two-dimensional square lattice as shown in Fig. 2. The lattice constant is taken as $a$. The length and width of QAHI are $N$ and $w$, respectively, which satisfy $L=(N-1)a$ and $W=(w-1)a$. The STS regions are of the same width but semi-infinite along the $x$ direction.

The discrete Hamiltonians for left STS and right STS are
\begin{eqnarray}
\begin{split}
H_{LS}=&\sum_{\substack{i_x\le0\\1\le i_y\le w}}\Psi^{L+}_{i}\check{H}^{L}_{0}\Psi^{L}_{i}
 +\sum_{\substack{i_x\le -1\\1\le i_y\le w}}
 \Psi^{L+}_{i}\check{H}^{L}_{x}\Psi^{L}_{i+\delta{x}}\\
&+\sum_{\substack{i_x\le 0\\1\le i_y\le w-1}} \Psi^{L+}_{i}\check{H}^{L}_{y}\Psi^{L}_{i+\delta{y}}+H.C.,\label{dhls}
\end{split}
\end{eqnarray}
and
\begin{eqnarray}
\begin{split}
H_{RS}=&\sum_{\substack{i_x\geq N+1\\1\le i_y\le w}}\left[\Psi^{R+}_{i}\check{H}^{R}_{0}\Psi^{R}_{i}
 +\Psi^{R+}_{i}\check{H}^{R}_{x}\Psi^{R}_{i+\delta{x}}\right]\\
&+\sum_{\substack{i_x\geq N+1\\1\le i_y\le w-1}} \Psi^{R+}_{i}\check{H}^{R}_{y}\Psi^{R}_{i+\delta{y}}+H.C.,\label{dhrs}
\end{split}
\end{eqnarray}
respectively. Here $\Psi^{L(R)}_{i}=(\Psi_{L(R)i\uparrow},\Psi_{L(R)i\downarrow},\Psi^{\dagger}_{L(R)i\uparrow},\Psi^{\dagger}_{L(R)i\downarrow})^{T}$ in which $\Psi_{L(R)i\alpha}$ is the annihilation operator of electron with spin $\alpha$ on the site $i=(i_x,i_y)$ in the left(right) STS. The matrices are
$\check{H}^{L(R)}_{0}=\text{diag}(\frac{2}{ma^2}-{\mu_{L(R)}},\frac{2}{ma^2}-{\mu_{L(R)}},-\frac{2}{ma^2}+{\mu_{L(R)}},
-\frac{2}{ma^2}+{\mu_{L(R)}})$,
\begin{eqnarray}
\check{H}^{L(R)}_{x}=\left(\begin{array}{cccc}
-\frac{1}{2ma^2}&0&\frac{-i\Delta^{L(R)}_{\uparrow\uparrow}}{2a}&\frac{-i\Delta^{L(R)}_{\uparrow\downarrow}}{2a}\\
0&-\frac{1}{2ma^2}&\frac{-i\Delta^{L(R)}_{\downarrow\uparrow}}{2a}&\frac{-i\Delta^{L(R)}_{\downarrow\downarrow}}{2a}\\
\frac{-i\Delta^{L(R)*}_{\uparrow\uparrow}}{2a}&\frac{-i\Delta^{L(R)*}_{\downarrow\uparrow}}{2a}&\frac{1}{2ma^2}&0\\
\frac{-i\Delta^{L(R)*}_{\uparrow\downarrow}}{2a}&\frac{-i\Delta^{L(R)*}_{\downarrow\downarrow}}{2a}&0&\frac{1}{2ma^2}
\end{array}\right),
\end{eqnarray}
and
\begin{eqnarray}
\check{H}^{L(R)}_{y}=\left(\begin{array}{cccc}
-\frac{1}{2ma^2}&0&\frac{\Delta^{L(R)}_{\uparrow\uparrow}}{2a}&\frac{\Delta^{L(R)}_{\uparrow\downarrow}}{2a}\\
0&-\frac{1}{2ma^2}&\frac{\Delta^{L(R)}_{\downarrow\uparrow}}{2a}&\frac{\Delta^{L(R)}_{\downarrow\downarrow}}{2a}\\
\frac{-\Delta^{L(R)*}_{\uparrow\uparrow}}{2a}&\frac{-\Delta^{L(R)*}_{\downarrow\uparrow}}{2a}&\frac{1}{2ma^2}&0\\
\frac{-\Delta^{L(R)*}_{\uparrow\downarrow}}{2a}&\frac{-\Delta^{L(R)*}_{\downarrow\downarrow}}{2a}&0&\frac{1}{2ma^2},
\end{array}\right),
\end{eqnarray}
where $\Delta^{L(R)}_{\uparrow\uparrow}=\Delta(-n_{L(R)1}+in_{L(R)2})e^{i\phi_{L(R)}}$, $\Delta^{L(R)}_{\uparrow\downarrow}
=\Delta^{L(R)}_{\downarrow\uparrow}=\Delta n_{L(R)3}e^{i\phi_{L(R)}}$ and
$\Delta^{L(R)}_{\downarrow\downarrow}=\Delta(n_{L(R)1}+in_{L(R)2})e^{i\phi_{L(R)}}$ with ${\bf{n}}_{L(R)}=(n_{L(R)1},n_{L(R)2},n_{L(R)3})$.

The discrete Hamiltonian for QAHI is
\begin{eqnarray}
\begin{split}
H_{QAHI}=&\sum_{\substack{1\le i_x\le N\\1\le i_y\le w}}\psi^{\dagger}_{i}\check{H}_{0}\psi_{i}
+\sum_{\substack{1\le i_x\le N-1\\1\le i_y\le w}} \psi^{\dagger}_{i}\check{H}_{x}\psi_{i+\delta x}\\
+&\sum_{\substack{1\le i_x\le N\\1\le i_y\le w-1}}\psi^{\dagger}_{i}\check{H}_{y}\psi_{i+\delta y}+H.C.,\label{LQ}
\end{split}
\end{eqnarray}
with $\psi_{i}=(\psi_{i\uparrow},\psi_{i\downarrow},\psi^{\dagger}_{i\uparrow},\psi^{\dagger}_{i\downarrow})^{T}$ in which $\psi_{i\alpha}$ is the annihilation operator of electron with spin $\alpha$ on the site $i=(i_x,i_y)$ in QAHI regime. The matrices are
$\check{H}_0=\text{diag}({m_0}+\frac{4B}{a^2},-{m_0}-\frac{4B}{a^2},-{m_0}-\frac{4B}{a^2},{m_0}+\frac{4B}{a^2})$,
\begin{eqnarray}
\check{H}_{x}=\left(\begin{array}{cccc}
-\frac{B}{a^2}&-\frac{iA}{2a}&0&0\\
-\frac{iA}{2a}&\frac{B}{2a^2}&0&0\\
0&0&\frac{B}{a^2}&-\frac{iA}{2a}\\
0&0&-\frac{iA}{2a}&-\frac{B}{a^2}
\end{array}\right),
\end{eqnarray}
and
\begin{eqnarray}
\check{H}_{y}=\left(\begin{array}{cccc}
-\frac{B}{a^2}&-\frac{A}{2a}&0&0\\
\frac{A}{2a}&\frac{B}{a^2}&0&0\\
0&0&\frac{B}{a^2}&\frac{A}{2a}\\
0&0&-\frac{A}{2a}&-\frac{B}{a^2}
\end{array}\right).
\end{eqnarray}
When we consider a QAHI ribbon with open boundary conditions at $y=0$ and $W$, $k_x$ will be a good quantum number. The energy bands of QAHI can be calculated from the lattice model in Eq. (\ref{LQ}), which has been shown in Fig. 1(b).

The tunneling Hamiltonian describing hopping between different regions can be written as
\begin{eqnarray}
H_{T}=\sum_{\substack{1\le i_y\le w}}\left[\Psi^{L+}_{0}\check{T}\psi_{1}
+\Psi^{R+}_{N+1}\check{T}\psi_{N}+H.C.\right],
\end{eqnarray}
with the hoping matrix $\check{T}=\text{diag}(t,t,-t^{*},-t^{*})$.
For simplicity, we use the subscript $0$ in $\Psi^{L+}_{0}$ to denote the site $i=(0,i_y)$.
The subscripts in other operators have the same meanings.

\subsection{Expression of Josephson current}
We define the particle number operator for the left STS as
\begin{eqnarray}
N_{L}=\sum_{\substack{i_x\le 0\\1\le i_y\le w}}\sum_{\alpha}\Psi^{+}_{Li\alpha}\Psi_{Li\alpha}.
\end{eqnarray}
The Josephson current is given by
\begin{eqnarray}
I=e\left\langle\frac{dN_{L}}{dt}\right\rangle=-e\sum_{i_{y}}\text{Tr}[\Gamma_{z}\check{T}G_{QS}^{<}(t,t,\substack{1\\i_{y}},\substack{0\\i_{y}})+H.C.],
\end{eqnarray}
with $\Gamma_z=\sigma_z\otimes 1_{2\times2}$. The ``lesser" Green function is defined as $G^{<}_{QS}(t,t',\substack{1\\i_{y}},\substack{0\\i'_{y}})=i\langle \Psi^{L+}_{(0,i'_y)}(t')\otimes\psi_{(1,i_y)}(t)\rangle$.

By introducing the contour-ordered Green function and using Langreth theorem, the current can be expressed as\cite{Sun,Li2,Song}
\begin{eqnarray}
\begin{split}
I&=-\frac{e}{2\pi}\int dE\text{Tr}[\Gamma_{z}G_{Q}^{r}(E)\Sigma_{LS}^{<}(E)+\Gamma_{z}G_{Q}^{<}(E)\Sigma_{LS}^{a}(E)\\
&-\Gamma_{z}\Sigma_{LS}^{<}(E)G_{Q}^{a}(E)-\Gamma_{z}\Sigma_{LS}^{r}(E)G_{Q}^{<}(E)].
\end{split}
\end{eqnarray}
Here, the Green functions $G_{Q}^{r}(E),G_{Q}^{a}(E)$ and $G_{Q}^{<}(E)$ in QAHI regime can be derived by the recursive algorithm. The self energies are given by $\Sigma_{LS}^{r}(E)=\check{T}^{\dagger}g_{LS}^{r}(E)\check{T}$, $\Sigma_{LS}^{a}(E)=\check{T}^{\dagger}[g_{LS}^{r}(E)]^{\dagger}\check{T}$ and $\Sigma_{LS}^{<}(E)=-f(E)[\Sigma_{LS}^{r}(E)-\Sigma_{LS}^{a}(E)]$ with $f(E)$ the Fermi distribution function. The surface Green function $g_{LS}^{r}(E)$ for the left STS can be deduced by the M$\ddot{\text{o}}$bius transformation according to Ref.[{\onlinecite{Umerski}}]. The detailed derivation for the Green functions is presented in Appendix.

\section{\label{sec3}Numerical results}
We will discuss two types of junction configurations, the parallel one and the nonparallel one. For the first case, the ${\bf{d}}$-vectors in the two STSs keep the same orientation and are rotated together. For the second case, the ${\bf{d}}$-vector in the left STS is fixed along the $z$-axis, i.e., ${\bf{d}_{L}\parallel \hat{z}}$, while the ${\bf{d}}$-vector for the right STS is rotated arbitrarily. In our calculations, we take $a=0.5$, $N=w=40$, $m=2$, $\mu_{L}=\mu_{R}=2.5$, $t=1$, $\Delta=0.005$
and the temperature $\mathcal{T}=0$.
The current almost keep the same value at the low temperature,
e.g. $\mathcal{T}<0.05\mathcal{T}_C$. The superconductor gap $\Delta$ is far less than the bulk gap $E_{g}$ of QAHI in our calculations. This ensures that the current flows only through the chiral edge states of QAHI. The realistic values of $\Delta$ and $E_{g}$ in experiment can well  meet the requirement $\Delta\ll E_{g}$.\cite{Mackenzie,Deng,Mogi,Lee}
The unit of the current is chosen as $\frac{e\Delta}{\pi}$. Since the Josephson current only depends on the phase difference, we will define $\phi=\phi_{L}-\phi_{R}$.

It is well known that the Josephson current can be generally decomposed into the Fourier series\cite{Tanaka}, $I(\phi)=\sum_{n\ge1}[a_{n}\sin{(n\phi)}+b_{n}\cos{(n\phi)}]$. Accordingly, the free energy of Josephson junctions can be given by
$E(\phi)=\frac{1}{2e}\sum_{n\ge1}[\frac{a_{n}}{n}(1-\cos{n\phi})+\frac{b_{n}}{n}\sin{n\phi}]$. For the junctions composed of a spin-singlet superconductor and a STS, the lowest order current with $n=1$ is absent due to the orthogonality of the wave functions of Cooper pairs\cite{Pals}. However, it is not the case for the STS-QAHI-STS junctions where the lowest order current usually exists. Therefore, one approximately has $I(\phi)=a_{1}\sin{\phi}+b_{1}\cos{\phi}$ and $E(\phi)=\frac{1}{2e}[a_{1}(1-\cos{\phi})+b_{1}\sin{\phi}]$. Since the Josephson current $I(\phi)$ is also a function of orientations of $\bf{d}$-vectors, we will express the current as $I(\theta_{L},\theta_{R},\varphi_{L},\varphi_{R},\phi)$ in the next sections.

\subsection{Parallel configuration}
\begin{figure}[!htb]
\centerline{\includegraphics[width=1\columnwidth]{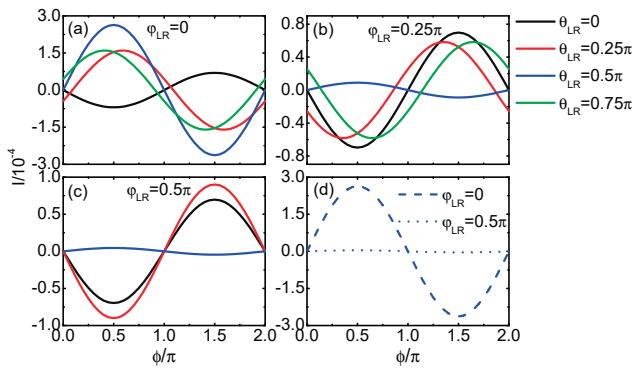}}
\caption{The CPRs for different polar angles of ${\bf{d}}_{L}$ and ${\bf{d}}_{R}$ at (a) $\varphi_{LR}=0$, (b) $\varphi_{LR}=0.25\pi$ and (c) $\varphi_{LR}=0.5\pi$. (d) The CPRs for $\theta_{LR}=0.5\pi$ at $\varphi_{LR}=0$(dashed line) and $0.5\pi$(dotted line) are plotted together for comparison. There are only three curves in (c) because the current for $\theta_{LR}=0.75\pi$ is the same as that for $\theta_{LR}=0.25\pi$ when $\varphi_{LR}=0.5\pi$.}
\label{fig1}
\end{figure}

For the parallel situation, we will use $\theta_{LR}$ and $\varphi_{LR}$ to denote the common polar angle and azimuthal angle of STSs for simplicity. In this situation, the current can be expressed simply as $I(\theta_{LR},\varphi_{LR},\phi)$ due to the relations $\theta_{L}=\theta_{R}=\theta_{LR}$ and $\varphi_{L}=\varphi_{R}=\varphi_{LR}$. Fig. 3 shows the dependence of CPRs on  orientations of ${\bf{d}}$-vectors. Fig. 3(a) gives CPRs for different polar angles at the azimuthal angle $\varphi_{LR}=0$. In this case, the ${\bf{d}}$-vectors are rotated in the $xz$ plane. It is found that CPRs are of the form $I(\phi)=a_{1}\sin{\phi}$ with $b_{1}=0$ when the ${\bf{d}}$-vectors are parallel to the $z$-axis or the $x$-axis, i.e., $\theta_{LR}=0$ or $0.5\pi$. Moreover, $a_{1}$ is positive for $\theta_{LR}=0.5\pi$ while it is negative for $\theta_{LR}=0$. For the former case, the free energy $E(\phi)$ achieves its minimum at $\phi=0$. The junctions are in the conventional $0$ phase. For the latter case, the minimum of $E(\phi)$ is obtained at $\phi=\pi$. So the $\pi$ phase can be realized in the junctions. In other words, the junctions can host the $0$-$\pi$ transition when one rotates the ${\bf{d}}$-vectors from the direction parallel to the $x$-axis to the direction parallel to the $z$-axis.
Additionally, the current for $\theta_{LR}=\pi$ is equal to that for $\theta_{LR}=0$. That is, the current is invariant when one inverses the ${\bf{d}}$-vectors from the $z$ direction to the $-z$ direction.

When the ${\bf{d}}$-vectors have both the $x$ and $z$ components, the $\cos{\phi}$-type current will emerge as shown in Fig. 3(a) for $\theta_{LR}=0.25\pi$ and $0.75\pi$. We have the CPRs of the form $(a_{1}\sin{\phi}+b_{1}\cos{\phi})$ with $a_{1}\ne0$ and $b_{1}\ne0$. In this situation, the phase difference for the free energy minimum is not at $\phi=0$ or $\pi$, but at $\phi=\phi_{0}$. The Josephson current no longer vanishes at the zero phase difference. For $\theta_{LR}=0.25\pi$, $\phi_{0}$ is between $0$ and $\pi$ while for $\theta_{LR}=0.75\pi$, $\phi_{0}$ is between $\pi$ and $2\pi$. Actually, the two current curves satisfy the following symmetry relation,
\begin{eqnarray}
I(\theta_{LR},\varphi_{LR},\phi)=-I(\pi-\theta_{LR},\varphi_{LR},-\phi).\label{sr1}\label{p1}
\end{eqnarray}

Fig. 3(b) shows the CPRs for different polar angles at $\varphi_{LR}=0.25\pi$. The $0$-$\pi$ transition still exists for rotation from $\theta_{LR}=0.5\pi$ to $\theta_{LR}=0$. When $\theta_{LR}$ deviates from the two values, the $\phi_{0}$ phase will be realized. The CPRs for $\theta_{LR}=0.25\pi$ and $0.75\pi$ also satisfy the symmetry relation presented in Eq. ({\ref{sr1}}).

Fig. 3(c) shows the CPRs for different polar angles at $\varphi_{LR}=0.5\pi$. Distinct from CPRs for $\varphi_{LR}=0$ and $0.25\pi$ given in Figs. 3(a) and (b), there is only $0$-$\pi$ transition and no $\phi_{0}$ phase is formed in this situation. This is because the $\cos{\phi}$-type current will disappear when ${\bf{d}}$-vectors are rotated in the $yz$ plane with $\varphi_{LR}=0.5\pi$. The same thing will happen when ${\bf{d}}$-vectors are rotated in the $xy$ plane with $\theta_{LR}=0.5\pi$ as shown in Figs. 3(a)-(c).
This indicates the necessary condition for the appearance of $\cos{\phi}$ term or the formation of the $\phi_{0}$ phase is
\begin{eqnarray}
d_{x}d_{z}\ne0.\label{SR}
\end{eqnarray}
It is reasonable to speculate $b_{1}\propto d_{x}d_{z}$ in $I(\phi)$ and $E(\phi)$. However, the existence of $\sin{\phi}$-type current is independent of the rotation of ${\bf{d}}$-vectors. The selection rule for the $\cos{\phi}$-type current and the $\phi_{0}$ phase in Eq. ({\ref{SR}}) is a peculiar feature for the STS-QAHI-STS junctions. It is meaningful to compare our results to those for the spin-singlet superconductor$-$QAHI$-$spin-singlet superconductor junctions in Ref.[{\onlinecite{Sakurai}}]. There, the formation of $\phi_{0}$ phase requires an extra Zeeman field or an asymmetric junction geometry.

From Figs. 3(a)-(c), we can also find that the current is dramatically weakened when ${\bf{d}}$-vectors are rotated from the direction along the $x$ axis to the direction along the $y$ axis in the $xy$ plane. This will become clear if we plot the curves for $(\varphi_{LR},\theta_{LR})=(0,0.5\pi)$ and $(\varphi_{LR},\theta_{LR})=(0.5\pi,0.5\pi)$ together as shown in Fig. 3(d). The huge current ration leads to on/off behavior of the Josephson current.

We do not show CPRs for $\varphi_{LR}$ with lager values, since the following symmetries hold for the junctions,
\begin{eqnarray}
I(\theta_{LR},\varphi_{LR},\phi)=I(\theta_{LR},2\pi-\varphi_{LR},\phi),\label{p2}
\end{eqnarray}
and
\begin{eqnarray}
I(\theta_{LR},\varphi_{LR},\phi)=I(\pi-\theta_{LR},\pi-\varphi_{LR},\phi).\label{p3}
\end{eqnarray}
The symmetries of CPRs in Eqs.(\ref{p1}),(\ref{p2}) and (\ref{p3}) possess direct correlations to the invariance obeyed by QAHI and we will discuss them later.

\subsection{Nonparallel configuration}
\begin{figure}[!htb]
\centerline{\includegraphics[width=1\columnwidth]{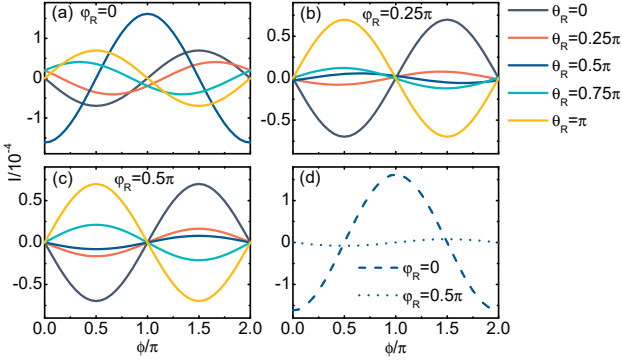}}
\caption{The CPRs for different polar angles of ${\bf{d}}_{R}$ at (a) $\varphi_{R}=0$, (b) $\varphi_{R}=0.25\pi$ and (c) $\varphi_{R}=0.5\pi$. (d) The CPRs for $\theta_{R}=0.5\pi$ at $\varphi_{R}=0$(dashed line) and $0.5\pi$(dotted line) are plotted together for comparison. The ${\bf{d}}_{L}$-vector is fixed along the $z$ axis.}
\label{fig1}
\end{figure}
For the nonparallel situation, the polar angle for ${\bf{d}}_{L}$ is taken as $\theta_{L}=0$. Fig. 4 shows the dependence of CPRs on the orientation of ${\bf{d}}_{R}$-vector. Fig. 4(a) gives the CPRs for different values of the polar angle $\theta_{R}$ at $\varphi_{R}=0$. The junctions host the $0$ phase at $\theta_{R}=\pi$ and the $\pi$ phase at $\theta_{R}=0$. The $0$-$\pi$ transition happens when one inverses the vector from the $-z$ direction to the $+z$ direction. As the ${\bf{d}}_{R}$-vector has the nonzero $x$ component such as $\theta_{R}=0.25\pi$ and $0.75\pi$, the $\cos{\phi}$ term in $I(\phi)$ appears. The $\phi_{0}$ phase will be achieved. Especially, the term $\cos{\phi}$ dominates the Josephson current when $\theta_{R}=0.5\pi$. The minimum of the free energy is obtained at $\phi_{0}\approx\frac{\pi}{2}$.
Fig. 4(b) gives the CPRs for different $\theta_{R}$ at $\varphi_{R}=0.25\pi$. The $0$-$\pi$ transition still exist in this case. However, the $\cos{\phi}$-type current is suppressed compared with CPRs in Fig. 4(a). The $\phi_{0}$ phase evolves towards the $0$ and $\pi$ phases as $\varphi_{R}$ is increased from $0$ to $0.25\pi$.

When $\varphi_{R}=0.5\pi$ as given in Fig. 4(c), the $\cos{\phi}$-type current disappears and the $\phi_{0}$ phase cannot be realized in the junctions. There are only the $0$-$\pi$ transitions. From Figs. 4(a)-(c), we can summarize that the necessary condition for $\cos{\phi}$-type current or the formation of the $\phi_{0}$ phase is that the x-component of the ${\bf d}$-vector in the
right STS is nonzero
\begin{eqnarray}
d_{Rx}\ne0,
\end{eqnarray}
when the ${\bf{d}_{L}}$-vector is fixed along the $z$-axis. In Fig. 4(d), we plot together the CPRs for $\varphi_{R}=0$ and $0.5\pi$ when the ${\bf{d}_{R}}$-vector lies in the $xy$ plane with $\theta_{R}=0.5\pi$. The current is dramatically weakened when one rotates the vector from the $x$ direction to the $y$ direction. In addition, $I(\phi)\approx b_{1}\cos{\phi}$ for $\varphi_{R}=0$ while $I(\phi)= a_{1}\sin{\phi}$ for $\varphi_{R}=0.5\pi$, hence the junctions can be used as a current switch at the fixed phase difference $0$ or $\pi$.

The CPRs $I(\theta_{L},\theta_{R},\varphi_{L},\varphi_{R},\phi)$ for the nonparallel configuration satisfy the following symmetry relations,
\begin{eqnarray}
\begin{split}
I(0,\theta_{R},\varphi_{L},\varphi_{R},\phi)&=-I(0,\theta_{R},\pi-\varphi_{L},\pi-\varphi_{R},-\phi),\\
I(0,\theta_{R},\varphi_{L},\varphi_{R},\phi)&=I(\pi,\pi-\theta_{R},\pi+\varphi_{L},\pi+\varphi_{R},\phi),\\
I(0,\theta_{R},\varphi_{L},\varphi_{R},\phi)&=I(\theta_{R},0,2\pi-\varphi_{R},2\pi-\varphi_{L},\phi).\label{np1}
\end{split}
\end{eqnarray}

Although the two configurations (the parallel one and the nonparallel one) are different, the relations satisfied by CPRS are consistent. For example, the combination of Eqs.(\ref{p1}) and (\ref{p3}) gives $I(\theta_{LR},\varphi_{LR},\phi)=-I(\theta_{LR},\pi-\varphi_{LR},-\phi)$ which is in line with the first equality in Eq.(\ref{np1}). This implies that the STS-QAHI-STS junctions have some universal symmetry relations of CPRS. We will present the their derivation in the next section through the symmetry analysis of Hamiltonians.

\section{\label{sec4}Symmetry analysis}
The behaviours of CPRs in Josephson junctions are closely related to Hamiltonians of junctions.\cite{Sakurai,Qiang,Qiang2}
Now, we derive the symmetries of CPRs from the continuum model in Eq. (\ref{HLQR}). We introduce five transformation operators: (1) the time-reversal $\mathcal{T}$,(2) the mirror-reflection about the $xz$ plane $\mathcal{M}_{xz}$, (3) the mirror-reflection about
the $yz$ plane $\mathcal{M}_{yz}$, (4) the spin rotation of $\pi$ about the $z$-axis $\mathcal{R}_{z}(\pi)$ and (5) the gauge transformation $\mathcal{U}_{1}(\eta)$. Their actions on the annihilation operators are given by
\begin{eqnarray}
\begin{split}
\mathcal{T}c_{\bf{k}\alpha}\mathcal{T}^{-1}&=\alpha c_{-\bf{k}\bar{\alpha}},\\
\mathcal{M}_{xz}c_{(k_{x},k_{y})\alpha}\mathcal{M}_{xz}^{-1}&=\alpha c_{(k_{x},-k_{y})\bar{\alpha}},\\
\mathcal{M}_{yz}c_{(k_{x},k_{y})\alpha}\mathcal{M}^{-1}_{yz}&=-ic_{(-k_{x},k_{y})\bar{\alpha}},\\
\mathcal{R}_{z}(\pi)c_{\bf{k}\alpha}\mathcal{R}^{-1}_{z}(\pi)&=\alpha ic_{\bf{k}\alpha},\\
\mathcal{U}_{1}(\eta)c_{\bf{k}\alpha}\mathcal{U}_{1}^{-1}(\eta)&=c_{\bf{k}\alpha}e^{i\eta},
\end{split}
\end{eqnarray}
with $\alpha(\bar{\alpha})=\uparrow\downarrow(\downarrow\uparrow)$ or $\pm(\mp)$. The matrices for the transformation operators
$\mathcal{T}$, $\mathcal{M}_{xz}$, $\mathcal{M}_{yz}$, $\mathcal{R}_{z}(\pi)$ and $\mathcal{U}_{1}(\eta)$ are shown in Appendix A.3.

First, the center finite QAHI is invariant under the joint transformation $\mathcal{X}=\mathcal{R}_{z}(\pi)\mathcal{T}\mathcal{M}_{xz}$, i.e.,
\begin{eqnarray}
\mathcal{X}H_{QAHI}\mathcal{X}^{-1}=H_{QAHI},
\end{eqnarray}
but the same transformation can change the Hamiltonians of STSs according to
\begin{eqnarray}
\begin{split}
\mathcal{X}H_{LS}(\theta_{L},\varphi_{L},\phi_{L})\mathcal{X}^{-1}&=H_{LS}(\pi-\theta_{L},2\pi-\varphi_{L},-\phi_{L}),\\
\mathcal{X}H_{RS}(\theta_{R},\varphi_{R},\phi_{R})\mathcal{X}^{-1}&=H_{RS}(\pi-\theta_{R},2\pi-\varphi_{R},-\phi_{R}).
\end{split}
\end{eqnarray}
Although the operations $\mathcal{R}_{z}(\pi)$ and $\mathcal{M}_{xz}$
do not alter the direction of the Josephson current,
the time-reversal operation can inverse the direction of the current. Therefore, we obtain the following relation,
\begin{eqnarray}
\begin{split}
&I(\theta_{L},\theta_{R},\varphi_{L},\varphi_{R},\phi)\\
&=-I(\pi-\theta_{L},\pi-\theta_{R},2\pi-\varphi_{L},2\pi-\varphi_{R},-\phi).\label{Isr1}
\end{split}
\end{eqnarray}

Secondly, the center finite QAHI is invariant under the joint transformation $\mathcal{Y}=\mathcal{R}_{z}(\pi)\mathcal{T}\mathcal{M}_{yz}$, i.e.,
\begin{eqnarray}
\mathcal{Y}H_{QAHI}\mathcal{Y}^{-1}=H_{QAHI},
\end{eqnarray}
but the same transformation can change the Hamiltonians of STSs according to
\begin{eqnarray}
\begin{split}
\mathcal{Y}H_{LS}(\theta_{L},\varphi_{L},\phi_{L})\mathcal{Y}^{-1}&=H_{RS}(\theta_{L},2\pi-\varphi_{L},-\phi_{L}),\\
\mathcal{Y}H_{RS}(\theta_{R},\varphi_{R},\phi_{R})\mathcal{Y}^{-1}&=H_{LS}(\theta_{R},2\pi-\varphi_{R},-\phi_{R}).
\end{split}
\end{eqnarray}
Because $\mathcal{M}_{yz}$ will alter the axis $x\rightarrow-x$, the current is reversed in the original coordinate system. After the
time-reversal operation $\mathcal{T}$, the current changes back to the original direction. Hence, we obtain
\begin{eqnarray}
\begin{split}
&I(\theta_{L},\theta_{R},\varphi_{L},\varphi_{R},\phi)\\
&=I(\theta_{R},\theta_{L},2\pi-\varphi_{R},2\pi-\varphi_{L},\phi).\label{Isr2}
\end{split}
\end{eqnarray}
Note, the polar angle and the azimuthal angle for the left STS and those for the right STS have been interchanged.

Thirdly, the center finite QAHI is invariant under the gauge transformation $U_{1}(\eta)$, i.e.,
\begin{eqnarray}
\mathcal{U}_{1}(\eta)H_{QAHI}\mathcal{U}_{1}^{-1}(\eta)=H_{QAHI}.
\end{eqnarray}
If one choose $\eta=\frac{\pi}{2}$, the Hamiltonians of STSs will be changed into
\begin{eqnarray}
\begin{split}
\mathcal{U}_{1}(\frac{\pi}{2})H_{LS}(\theta_{L},\varphi_{L},\phi_{L})\mathcal{U}_{1}^{-1}(\frac{\pi}{2})
&=H_{LS}(\pi-\theta_{L},\pi+\varphi_{L},\phi_{L}),\\
\mathcal{U}_{1}(\frac{\pi}{2})H_{RS}(\theta_{R},\varphi_{R},\phi_{R})\mathcal{U}_{1}^{-1}(\frac{\pi}{2})
&=H_{RS}(\pi-\theta_{R},\pi+\varphi_{R},\phi_{R}).
\end{split}
\end{eqnarray}
Since the unitary operation $\mathcal{U}_{1}(\eta)$ does not change the Josephson current, we can conclude the following symmetry relation
\begin{eqnarray}
\begin{split}
&I(\theta_{L},\theta_{R},\varphi_{L},\varphi_{R},\phi)\\
&=I(\pi-\theta_{L},\pi-\theta_{R},\pi+\varphi_{L},\pi+\varphi_{R},\phi).\label{Isr3}
\end{split}
\end{eqnarray}

One can easily prove that the derived symmetry relations of CPRs here from the invariance of $H_{QAHI}$ can immediately lead to the equalities in Eqs. (\ref{p1})-(\ref{p3}) and (\ref{np1}) summarized from numerical calculations for the parallel and nonparallel configurations. In addition, from Eqs.(\ref{Isr1}) and (\ref{Isr3}), we can find $I(\varphi_{L},\varphi_{R},\phi)=-I(\pi-\varphi_{L},\pi-\varphi_{R},-\phi)$ which is irrespective of $\theta_{L}$ and $\theta_{R}$. For $\varphi_{L}=\varphi_{R}=0.5\pi$, the relation means the pure $\sin{\phi}$ CPRs which are demonstrated in Figs.3(c) and 4(c). The deviation from $\varphi_{L}=\varphi_{R}=0.5\pi$ will ruins the pure $\sin{\phi}$ CPRs and causes the formation of the $\phi_{0}$ phase. From Eqs.(\ref{Isr1}) and (\ref{Isr2}), we can deduce $I(\theta_{LR},\phi)=-I(\pi-\theta_{LR},-\phi)$ irrespective of $\varphi_{LR}$ for the parallel configuration (see also Eq.(\ref{p1})). For $\theta_{L}=\theta_{R}=0.5\pi$, the relation also leads to the pure $\sin{\phi}$ CPRs as shown in Fig. 3. The deviation from $\theta_{L}=\theta_{R}=0.5\pi$ will break the pure $\sin{\phi}$ CPRs and the $\phi_{0}$ phase will form.

For spin-singlet superconductor$-$QAHI$-$spin-singlet superconductor junctions\cite{Sakurai}, the breaking of magnetic mirror reflection (the joint operation of the time reversal $\mathcal{T}$ and the mirror-reflection $\mathcal{M}_{xz}$)
symmetry is essential to form the $\phi_{0}$ phase. It can be achieved by exerting an extra field along the $y$-axis or constructing an asymmetric junctions with different width of superconductors and QAHIs. However, for the STS-QAHI-STS junctions here, the $\phi_{0}$ phase can be realized through rotating $\bf{d}$-vectors to deviate from specific angles. It's also important to note that the $0$-$\pi$ transition in STS-QAHI-STS junctions can not be achieved in the spin-singlet case. These critical differences originate from the peculiar coupling of STS and QAHI.

Finally, we give some discussions of the size dependence of CPRs. The
Josephson currents show strong dependence on the width $w$ of the junctions.
In addition, the Josephson currents also depend on the length $N$ of QAHI. However, the size dependence of CPRs will not change our essential results including the symmetry relations of CPRs and the selection rules for the $\phi_{0}$ phase. The $0$-$\pi$ transition and the switch effect still exist in junctions with different values of the width $w$ and length $N$.

\section{\label{sec5}Conclusions}
We study CPRs in the STS-QAHI-STS Josephson junctions by the lattice nonequilibrium Green function theory. The junctions host rich physics due to the presence of ${\bf{d}}$-vectors in STSs and the unique electric structure of QAHI. The CPRs are strongly dependent on the directions of the two ${\bf{d}}$-vectors in STSs. The dependences are detailedly investigated for the parallel and the nonparallel case. The $0$-$\pi$ transitions, the $\phi_{0}$ phase and the current switch effects are found in the both situations. The selection rules for the $\cos{\phi}$-type current which is the essential element for the $\phi_{0}$ phase, are summarized from the numerical results. The CPRs satisfy three kinds of different symmetry relations, which are closely related to the selection rules. We analyse the origin of these relations through the invariance of QAHI and the changes of STSs under the operations of the time-reversal, mirror-reflections, the spin-rotation and the gauge transformation. Our results exhibit a new type of Josephson coupling based on STSs and QAHI, which provide helpful $0$-$\pi$ transition, $\phi_{0}$ phase and on/off effects for the device design. The strong dependence of CPRs on the ${\bf{d}}$-vector orientation may be used to detect the information of the spin-triplet paring in STSs.

\section*{\label{sec5}ACKNOWLEDGMENTS}
This work was financially supported by National Key R and D Program of China (2017YFA0303301),
NSF-China under Grants Nos. 11921005 and 11447175,
the Strategic Priority Research Program of Chinese Academy of Sciences (XDB28000000), and
the Natural Science Foundation of Shandong Province under Grants No. ZR2017QA009.

\section{Appendix}
\setcounter{equation}{0}
\setcounter{subsection}{0}
\renewcommand{\theequation}{A.\arabic{equation}}
\renewcommand{\thesubsection}{A.\arabic{subsection}}
\subsection{Surface Green functions for STSs}
STSs have been discretized into a series of slices as shown in Fig. 2. Each slice consists of $w$ lattice points. We define the Hamiltonian of an isolated slice as $H_{L(R)11}$ for the left(right) STS. The hopping Hamiltonian from one slice to its right neighbor slice is denoted by
$H_{L(R)12}$. The elements of $H_{L(R)11}$ and $H_{L(R)12}$ can be determined by the lattice model for STSs in Eqs. (\ref{dhls}) and (\ref{dhrs}) of the main text. Construct the M$\ddot{\text{o}}$bius transformation matrix\cite{Umerski}
\begin{eqnarray}
X_{L}=\left(\begin{array}{cc}
0&H^{-1}_{L12}\\
-H^{\dagger}_{L12}&[(E+i\gamma)-H_{L11}]H^{-1}_{L12}
\end{array}\right).
\end{eqnarray}
with $\gamma$ a small positive quantity.
It can be diagonalized as $U^{-1}_{L}X_{L}U_{L}=\text{diag}(\lambda_{L1},\lambda_{L2},\lambda_{L3},\cdots)$ with the eigenvalues satisfying $\vert\lambda_{L1}\vert<\vert\lambda_{L2}\vert<\vert\lambda_{L3}\vert<\cdots$. We assume the matrix $U_{L}$ has the following form
\begin{eqnarray}
U_{L}=\left(\begin{array}{cc}
U_{L11}&U_{L12}\\
U_{L21}&U_{L22}
\end{array}\right).
\end{eqnarray}
Then, the surface Green function for the left STS is given by $g^{r}_{LS}(E)=U_{L12}U^{-1}_{L22}$.

For the right STS, the M$\ddot{\text{o}}$bius transformation matrix is constructed as
\begin{eqnarray}
X_{R}=\left(\begin{array}{cc}
0&(H^{\dagger}_{R12})^{-1}\\
-H_{R12}&[(E+i\gamma)-H_{R11}](H^{\dagger}_{R12})^{-1}
\end{array}\right).
\end{eqnarray}
It can be diagonalized by $U_{R}$ in a similar way. The surface Green function for the right STS is given by $g^{r}_{RS}(E)=U_{R12}U^{-1}_{R22}$. With $g^{L}_{RS}(E)$ and $g^{R}_{RS}(E)$, the self energies in the main text will be obtained.

\subsection{Green functions for QAHI}
We denote the Hamiltonian for an isolated slice of QAHI as $H_{Q11}$ and the hopping Hamiltonian from one splice to its right neighbor slice as $H_{Q12}$. The Green function for the rightmost slice is
\begin{eqnarray}
G^{Rr}_{Q}(E,N)=[E-H_{Q11}-\tilde{T}g^{r}_{RS}(E)\tilde{T}^{\dagger}]^{-1},
\end{eqnarray}
with $\tilde{T}=1_{w\times w}\otimes \check{T}$. The $n$th slice Green function can be derived from the following recursive algorithm,
\begin{eqnarray}
G^{Rr}_{Q}(E,n)=[E-H_{Q11}-H_{Q12}G^{Rr}_{Q}(E,n+1)H_{Q21}]^{-1}.
\end{eqnarray}
The full retarded Green function for the leftmost slice is given by
\begin{eqnarray}
\begin{split}
G^{r}_{Q}(E)&=[E-H_{Q11}-\tilde{T}g^{r}_{LS}(E)\tilde{T}^{\dagger}\\
&-H_{Q12}G^{Rr}(E,2)H_{Q21}]^{-1}.
\end{split}
\end{eqnarray}
The full advanced Green function is obtained by the relation $G^{a}_{Q}(E)=[G^{r}_{Q}(E)]^{\dagger}$.
Then, the full ``lesser" Green function for the leftmost slice of QAHI can be written as
\begin{eqnarray}
G_{Q}^{<}(E)=-f(E)(G_{Q}^{r}(E)-G_{Q}^{a}(E)).
\end{eqnarray}
With $G^{r}_{Q}(E)$ and $G_{Q}^{<}(E)$, the Josephson current can be calculated numerically.
\subsection{Matrices for transformation operators}
Here, we present the transformation matrices for five operators introduced in the main text. The matrix for the time-reversal operator is given by
\begin{eqnarray}
U_{\mathcal{T}}=\left(\begin{array}{cc}
-i\sigma_{y}&0\\
0&-i\sigma_{y}
\end{array}\right)\mathcal{K},
\end{eqnarray}
with $\mathcal{K}$ being the complex conjugation operator. The matrix for the mirror-reflection about the $xz$ plane is
\begin{eqnarray}
U_{\mathcal{M}_{xz}}=
\left(\begin{array}{cc}
i\sigma_{y}&0\\
0&i\sigma_{y}
\end{array}\right)\mathcal{R}_{y},
\end{eqnarray}
with $\mathcal{R}_{y}$ being the reflection operator in the real space, which will lead to $y\rightarrow-y$ and $k_{y}\rightarrow-k_{y}$. The matrix for the mirror-reflection about the $yz$ plane is
\begin{eqnarray}
U_{\mathcal{M}_{yz}}=
\left(\begin{array}{cc}
i\sigma_{x}&0\\
0&-i\sigma_{x}
\end{array}\right)\mathcal{R}_{x},
\end{eqnarray}
with $\mathcal{R}_{x}$ being the reflection operator in the real space, which will lead to $x\rightarrow-x$ and $k_{x}\rightarrow-k_{x}$.
The matrix for the spin rotation of $\pi$ angle about the $z$ axis is
\begin{eqnarray}
U_{\mathcal{R}_{z}(\pi)}=
\left(\begin{array}{cc}
-i\sigma_{z}&0\\
0&i\sigma_{z}
\end{array}\right).
\end{eqnarray}
The matrix for the gauge transformation $\mathcal{U}(\eta)$ is
\begin{eqnarray}
U_{\mathcal{U}(\eta)}=
\left(\begin{array}{cc}
e^{i\eta}1_{2\times2}&0\\
0&e^{-i\eta}1_{2\times2}
\end{array}\right),
\end{eqnarray}
with the identity matrix $1_{2\times2}$.

\end{document}